\documentclass[conference]{IEEEtran}
\IEEEoverridecommandlockouts
\makeatletter
\let\NAT@parse\undefined
\makeatother
\usepackage{amsmath}
\usepackage{hyperref}  %hyperref still needs to be put at the end!
 \usepackage{float}
\usepackage{cite}
\usepackage{amsmath,amssymb,amsfonts}
\usepackage{algorithmic}
\usepackage{graphicx}
\usepackage{textcomp}
\usepackage{xcolor}
\def\BibTeX{{\rm B\kern-.05em{\sc i\kern-.025em b}\kern-.08em
    T\kern-.1667em\lower.7ex\hbox{E}\kern-.125emX}}
\begin{document}

\title{Orthogonal Time Frequency Space with Delay-Doppler Alignment Modulation\\
% {\footnotesize \textsuperscript{*}N
}

\author{
	\IEEEauthorblockN{
		Xianda Liu\IEEEauthorrefmark{1}, 
		Zhiwen Zhou\IEEEauthorrefmark{1},
        Zhiqiang Xiao\IEEEauthorrefmark{1}\IEEEauthorrefmark{2}, and
        Yong Zeng\IEEEauthorrefmark{1}\IEEEauthorrefmark{2}}
	\IEEEauthorblockA{\IEEEauthorrefmark{1}National Mobile Communications Research Laboratory, Southeast University, Nanjing 210096, China}
	\IEEEauthorblockA{\IEEEauthorrefmark{2}Purple Mountain Laboratories, Nanjing 211111, China\\Email: \{220230853,220220790,zhiqiang\_xiao,yong\_zeng\}@seu.edu.cn.}
}
\maketitle
\begin{abstract}
Delay-Doppler alignment modulation (DDAM) is a novel technique to mitigate time-frequency doubly selective channels by leveraging the high spatial resolution offered by large antenna arrays and multi-path sparsity of millimeter wave (mmWave) and TeraHertz (THz) channels. By introducing per-path delay and Doppler compensations, followed by path-based beamforming, it is possible to reshape the channel features with significantly reduced channel delay and Doppler spreads. This offers new degrees-of-freedom for waveform designs such as orthogonal time frequency space (OTFS), since the reshaped channel can significantly relax the constraints on OTFS parameter selection and reduce the complexity of signal detection at the receiver. Therefore, in this paper, by combing DDAM with OTFS, we propose a novel technique termed DDAM-OTFS. Two implementation schemes are introduced for DDAM-OTFS, namely path-based alignment and bin-based alignment. Simulation results are provided to demonstrate the superior performance of the proposed DDAM-OTFS in terms of spectral efficiency (SE) and peak-to-average power ratio (PAPR) compared to the conventional OTFS.
\end{abstract}

% \begin{IEEEkeywords}
% Delay-Doppler alignment modulation, delayDoppler compensation, beamforming, OTFS
% \end{IEEEkeywords}

\section{Introduction}
In contemporary mobile communication systems, orthogonal frequency division multiplexing (OFDM) is the mainstream waveform. OFDM converts high-speed serial data into multiple low-speed parallel data streams, leading to OFDM symbols with the duration much longer than channel delay spread. Thus, it can effectively mitigate the inter-symbol interference (ISI) issue. However, for high mobility scenarios, the severe Doppler spread may cause inter-carrier interference (ICI), which greatly degrades the performance of OFDM. To address this issue, one possible approach is to shorten the OFDM symbol duration so that the channel variations over each symbol becomes inconsequential \cite{b1}. However, this inevitably increases the cyclic prefix (CP) overhead, consequently compromising spectral efficiency (SE). 

Orthogonal time frequency space (OTFS) is a recently proposed modulation scheme for doubly selective channels \cite{b2}, \cite{b369}. Compared to OFDM that modulates information in the time-frequency domain, OTFS modulates information symbols in the delay-Doppler domain. In scenarios with quasi-static and sparse time-varying channels, OTFS exhibits superior communication performance in high mobility scenarios. The extensions of OTFS for the multiple-input multiple-output (MIMO) case was presented in \cite{b102}, \cite{b192}. However, as the signal processing for OTFS is performed per frame basis, it often requires long signal processing delay and complex equalization at the receiver.

On the other hand, \textit{delay alignment modulation} (DAM) is a recently proposed technique that can effectively mitigate ISI by leveraging the high spatial resolution of large antenna arrays and the multi-path sparsity of millimeter-wave (mmWave) and TeraHertz (THz) channels \cite{b4}. By performing path-based delay compensation and beamforming at the transmitter and/or receiver, DAM is able to reduce the channel delay spread without sacrificing time-frequency resources. The extension of DAM to multi-user systems is studied in\cite{b1000}. In \cite{b5}, the authors proposed a DAM-OFDM scheme to jointly exploiting the advantages of DAM and OFDM for mitigating the ISI, where a unified single- and multi-carrier transmission scheme was proposed, which can significantly improve the performance in terms of SE, bit error rate (BER), and peak-to-average power ratio (PAPR). The application of DAM for integrated sensing and communication (ISAC) systems was investigated in \cite{b250}. In \cite{b248}, the fractional delay issue for DAM was studied for the practical implementation systems. Moreover, hybrid-beamforming-based DAM was studied in \cite{b249}. In \cite{b699}, the authors investigated the active reconfigurable intelligent surface (RIS)-assisted THz ISAC system with DAM. To address time-frequency doubly selective channels, delay-Doppler alignment modulation (DDAM) has been proposed \cite{b6}, which introduces additional path-based Doppler
compensation to manipulate Doppler spread as well.

In this paper, by combining the benefits of OTFS and DDAM, we propose a technique called DDAM-OTFS. Specifically, for OTFS, to avoid non-predictability and fading resulting from aliasing in the delay-Doppler domain, the selected delay and Doppler periods should be no smaller than the channel delay and Doppler spreads \cite{b3}, \cite{b103}, respectively. Under such constraints, it will become infeasible for OTFS to achieve certain specific goals in terms of SE and PAPR when the channel delay and Doppler spreads are large. The DDAM technique can significantly reduce the channel delay and Doppler spreads. Thus, combing DDAM with OTFS can relax the constraints on OTFS parameter selection, rendering enhanced performance on SE and PAPR. In this paper, two implementation methods of DDAM-OTFS are introduced for different situations, namely path-based alignment and bin-based alignment. Simulation results are provided to show the superiority of the proposed DDAM-OTFS in terms of the achievable SE and PAPR, compared to the conventional OTFS.

% \begin{itemize}
% \item Firstly, we discuss the constraints of  delay spread and Doppler spread on the selection of OTFS parameters. We show that DDAM can effectively reduce the delay and Doppler spread of the channel, maintaining the quasi-static nature of the channel. Combining DDAM with OTFS can relax these constraints and achieve better performance.
% \item Secondly, two implementation schemes of DDAM-OTFS are provided to address the varying difficulty of obtaining channel information in practical communication. These schemes include the path-based alignment scheme and the bin-based alignment scheme based on the nearest physical scattering radius. The signal-to-noise ratio (SNR) performance of the two schemes is derived.
% \item Lastly, to illustrate the performance improvement brought by DDAM-OTFS, the performance of DDAM-OTFS and OTFS is compared through simulation. The results show that due to the reduction of channel delay spread and Doppler spread by DDAM, DDAM-OTFS has significantly improved average SE and error rate performance.
% \end{itemize}

% The remaining sections of this article are organized as follows. Section\ref{section:A} briefly outlines the models of DDAM and OTFS. Section\ref{section:B} proposes a new communication scheme, called DDAM-OTFS, and introduces the input-output relationships and communication performance, including signal-to-noise ratio, of the two implementation schemes. Simulation results are presented in Section\ref{section:C}, and conclusion is drawn in Section\ref{section:D}.

\section{Motivation and System Model }
\label{section:A} %这里是章节的标签，引用时需要
\subsection{Motivation: Constraints on OTFS Parameters}
% In this section, we first review the main ideas of OTFS and DDAM. It is also pointed out that due to the constraints imposed by meeting the crystallization condition to ensure the non-fading property of the delay-Doppler domain, the SE and peak-to-average ratio performance of OTFS will be limited. This limitation is also the original intention behind introducing DDAM technology.
% \subsection{OTFS Concepts and Implementation}

 OTFS modulates symbols in the delay-Doppler domain. The information-bearing symbols ${X_{DD}}[k,l]$ in the delay-Doppler domain are transformed into the time domain signal ${x_{TD}}(t)$ through certain transformation methods, such as a combination of the symplectic finite Fourier transform (SFFT) and the inverse discrete Fourier transform (IDFT) or the inverse discrete Zak transform (IDZT) \cite{b7}. At the receiver side, the signal is processed through the corresponding inverse transformations to obtain the delay-Doppler domain signal ${Y_{DD}}[k,l]$. Further processing, such as equalization, can be applied to the received signal thereafter.

For the same time-domain signal, its representation in the delay-Doppler domain is non-unique. This mainly depends on the selection of the OTFS period parameters ${\tau _r}$ and ${\nu _r}$, which should satisfy ${\tau _r} \cdot {\nu _r} = 1$ \cite{b11}. Moreover, to avoid the delay-Doppler domain aliasing, the selection of the period parameters also needs to satisfy that $\tau_r >{\tau _{spread}}$ and $\nu_r >{\nu _{spread}} $, where ${\tau _{spread}}$ and ${\nu _{spread}}$ are the delay and Doppler spreads of the channel, respectively \cite{b3}. 
% They satisfy the following relationship: ${\tau _r} \cdot {\nu _r} = 1
%  ({\tau _r} = M\Delta \tau  = \frac{1}{{\Delta f}},{\nu _r} = N\Delta \nu  = \frac{1}{T})$\cite{b11}.
% \begin{equation}
% {\tau _r} \cdot {\nu _r} = 1
%  ({\tau _r} = M\Delta \tau  = \frac{1}{{\Delta f}},{\nu _r} = N\Delta \nu  = \frac{1}{T}),
% \label{eq10}
% \end{equation}

% Values are taken on a hyperbola as shown in Fig.\ref{fig2}.
% \begin{figure}[!t]
% \centerline{\includegraphics[width=0.4\textwidth]{p2.png}}
% \caption{ Delay-Doppler parametric representation.}
% \label{fig2}
% \end{figure}
  
  Additionally, requirements on SE and PAPR will also impose constraints on the selection of  ${\tau _r}$ and ${\nu _r}$. There are two ways to insert CP in OTFS \cite{b101}. One method is to insert one CP per time slot, while the other is to insert one CP per OTFS frame. The former method is similar to the traditional OFDM, which can effectively combat ISI. This method allows the receiver to mitigate the effects of channel Doppler shifts during signal detection and equalization, thereby reducing the complexity of signal detection at the receiver. On the other hand, adding one CP per OTFS frame can reduce CP overhead, but significantly increases the complexity of signal detection at the receiver \cite{b211}. So the introduction of DDAM may bring significant benefits for the first method. Denote the system bandwidth as ${B}$, the OTFS frame duration as ${T_{\rm{OTFS}}}$, and the sampling interval as $T = 1/B$. Consider an OTFS frame consisting of \textit{M} subcarriers and \textit{N} time slots, with subcarrier spacing $\Delta f = \frac{{{B}}}{M}$ and slot duration $T_s = \frac{{{T_{\rm{OTFS}}}}}{N}$. The OTFS period parameters ${\tau _r}$ and ${\nu _r}$ are ${\nu _r} = \Delta f$ and ${\tau _r} = {T_s}$. The CP overhead $\rho$ for one CP per time slot is 
\begin{equation}
\rho {\rm{ = }}\frac{{{n_{{\rm{CP}}}}}}{{M + {n_{{\rm{CP}}}}}} = \frac{{B{\tau _{spread}}}}{{M + B{\tau _{spread}}}},
\label{eq785}
\end{equation} 
   where ${n_{{\rm{CP}}}}$ is the CP length. Assuming the maximum tolerable CP overhead is ${\rho _{\max }}$, we have 
\begin{equation}
\frac{{B{\tau _{spread}}}}{{M + B{\tau _{spread}}}} \leq {\rho _{\max }} \Rightarrow B{\tau _{spread}}\frac{{1 - {\rho _{\max }}}}{{{\rho _{\max }}}} \leq M. 
\label{eq786}
\end{equation}    
Since $M = B/\Delta f = B \cdot {\tau _r}$, it has
\begin{equation}
\frac{{1 - {\rho _{\max }}}}{{{\rho _{\max }}}}{\tau _{spread}} \leq {\tau _r} 
\label{eq12}
\end{equation} 
% , as shown in Fig.\ref{fig4}.
% \begin{figure}[!t]
% \centerline{\includegraphics[width=0.4\textwidth]{DDAM-OTFS-2 (1).pdf}}
% \caption{ The limitations of CP on OTFS parameters.}
% \label{fig4}
% \end{figure}

Different from OFDM whose PAPR is proportional to the number of subcarriers, the PAPR of OTFS is related to the number of time slots \textit{N} in an OTFS frame. Assuming there is a maximum tolerable PAPR associated with maximum number of time slots ${N_{\max} }$: $N \leq {N_{\max }}$. Since $N = {T_{{\rm{OTFS}}}}/{T_s} = {T_{{\rm{OTFS}}}}/{\tau _r}$, the constraint can be rewritten
\begin{equation}
\frac{{{T_{\rm{OTFS}}}}}{{{N_{\max }}}} \leq {\tau _r}
\label{eq13}
\end{equation} 
In summary, we have the constraints on OTFS period parameters as follows
\begin{align}\left\{\begin{aligned}
&{{\tau _{spread}} < {\tau _r} < \frac{1}{{{\nu _{spread}}}}}\\
&{\frac{{1 - {\rho _{\max }}}}{{{\rho _{\max }}}}{\tau _{spread}} \le {\tau _r}}\\
&{\frac{{{T_{{\rm{OTFS}}}}}}{{{N_{\max }}}} \le {\tau _r}}
\label{eq1112}
\end{aligned}\right.\end{align}

% \begin{equation}
% \left\{ {\begin{array}{*{20}{c}}
% {{\tau _{spread}} < {\tau _r} < \frac{1}{{{\nu _{spread}}}}}\\
% {\frac{{1 - {\rho _{\max }}}}{{{\rho _{\max }}}}{\tau _{spread}} \le {\tau _r}}\\
% {\frac{{{T_{{\rm{OTFS}}}}}}{{{N_{\max }}}} \le {\tau _r}}
% \end{array}} \right,
% \label{eq13}
% \end{equation} 

\begin{figure}[H]
\centerline{\includegraphics[width=0.5\textwidth]{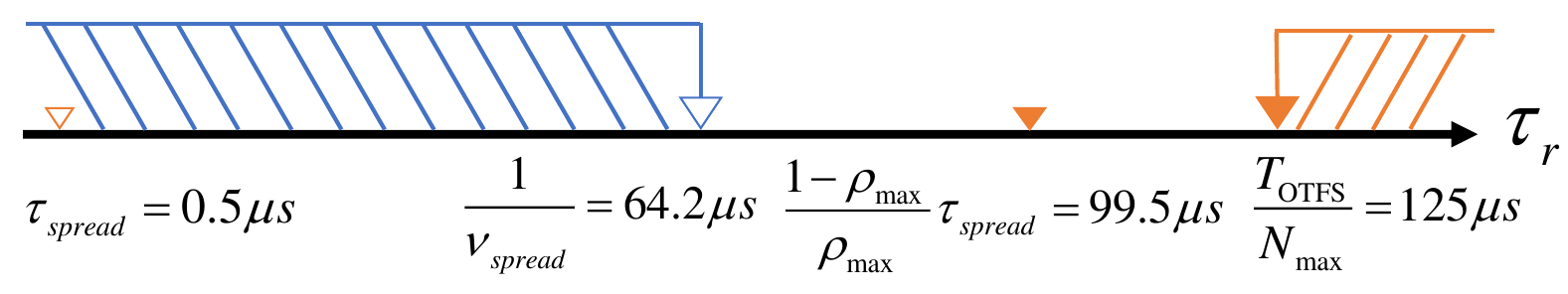}}
\caption{ The constraints on OTFS period parameters.}
\label{fig6}
\end{figure}

However, the above constraints may be infeasible for some practical scenarios when the channel delay Doppler spreads are large and/or when the targeting CP overhead $\rho_{max}$ and time slots $N_{max}$ are small. Here is a specific example to illustrate it. Assume the total bandwidth of the system is ${B} = 128\,MHz$, the duration of an OTFS frame is ${T_{\rm{OTFS}}} = 1\,ms$, the carrier frequency is ${f_c} = 28\,GHz$, ${\rho _{\max }}$ is 0.5\%, and ${N_{\rm{max }}}$ is 8, the maximum relative speed between the base station and the user is ${v_c}= 300\,km/h$, resulting in a maximum Doppler shift of $\frac{{{f_c}{v_c}}}{c} =  \pm 7.785\,kHz$. With Doppler spread of ${\nu _{spread}} = 15.57\,kHz$, delay spread of 500 ns, the following relationship must be satisfied ${125\,\mu s} \leq {\tau _r}{\rm{  < 64}}{\rm{.2}}\,\mu {\rm{s}}$, which is obviously infeasible. Fig. \ref{fig6} depicts the various constraint relationships. This issue motivates us to introduce DDAM to reduce channel delay and Doppler spreads, thereby relaxing the constraints and finding a feasible solution to the inequalities (\ref{eq1112}).
\subsection{System Model of DDAM-OTFS}
\label{section:B}
\begin{figure*}[!t]
\centerline{\includegraphics[width=0.9\textwidth]{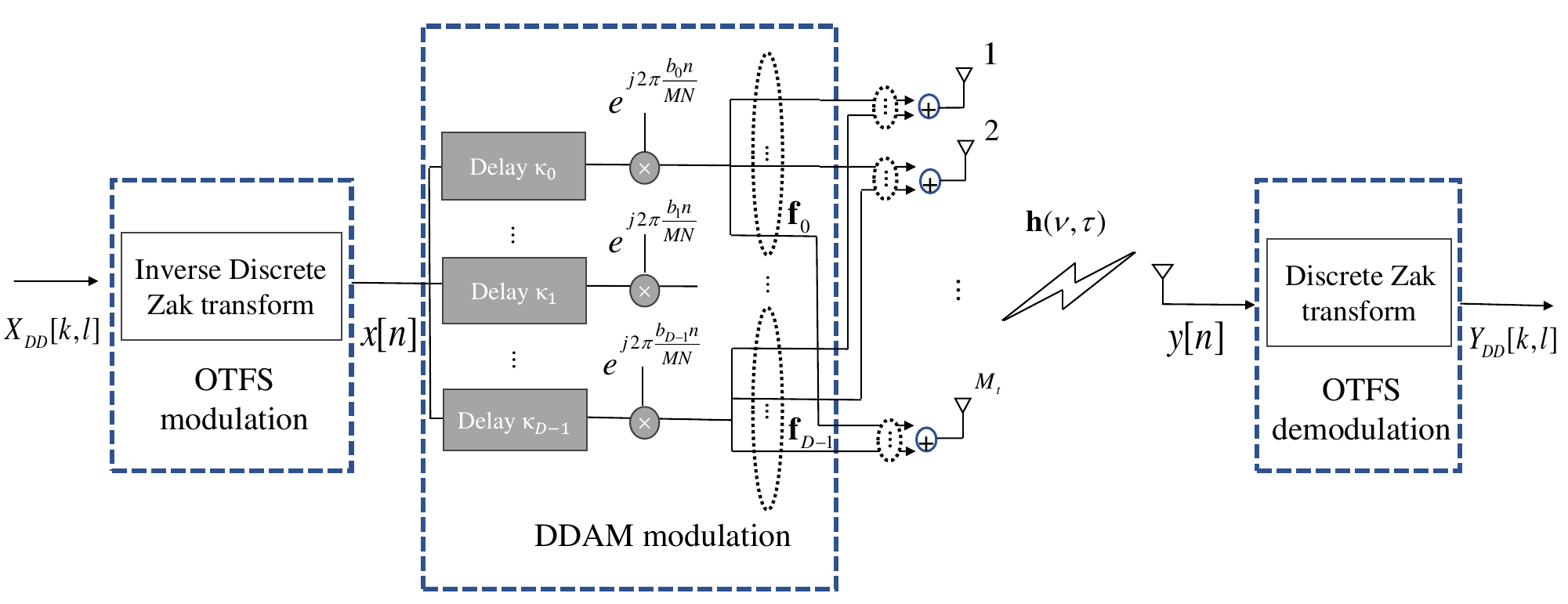}}
\caption{ The proposed DDAM-OTFS modulation.}
\label{fig8}
\end{figure*}

Fig. \ref{fig8} depicts the block diagram of the proposed DDAM-OTFS.
Each OTFS frame can multiplex a total of \textit{NM} quadrature amplitude modulation (QAM) symbols $\{ {X_{DD}}[k,l], k=0,...,N-1, l=0,...,M-1\} $, arranged on a delay-Doppler grid $\Gamma =\left\{ {\left( {\frac{k}{{NT_s}},\frac{l}{{M\Delta f}}} \right),k=0,...,N-1,l=0,...,M-1} \right\}$, where grid sizes on the delay and Doppler axis are $\Delta \tau  = \frac{1}{{M\Delta f}}$ and $\Delta \nu  = \frac{1}{{NT_s}}$, respectively. These two-dimensional data symbols are then sent to the OTFS modulator, where they undergo an IDZT transform to convert into one-dimensional time-domain symbols ${x}[n]$, which are then processed by DDAM \cite{b6}. To this end, where delay and Doppler compensations are applied before beamforming through $M_{t}$ transmit antennas. Upon receiving the signal, the receiver can directly perform OTFS demodulation, transforming the signal back to delay-Doppler domain symbols ${Y_{DD}}[k,l]$.

Let's discuss the specific input-output relationship. Performing IDZT on an OTFS frame \cite{b109} gives
\begin{equation}
{x}[n] = \frac{1}{{\sqrt N }}\sum\limits_{k = 0}^{N - 1} {{X_{DD}}[k,l]} {e^{  j2\pi \frac{c}{N}k}}{|_{n = l + cM}},
\end{equation}
where $l \in \left\{ {0,...,M - 1} \right\}$ and $c \in \left\{ {0,...,N - 1} \right\}$. Then, the time-domain signal undergoes DDAM processing, including \textit{D} sets of delay-Doppler compensations and beamformings \cite{b6}
\begin{equation}
{\bf{s}}[n] = \sum\limits_{d = 0}^{D-1} {{{\bf{f}}_d}x[n - {\kappa _d}]{e^{j2\pi \frac{{{b_d}n }}{{MN}}}}}  , 
\label{eq16}
\end{equation}
where ${{\bf{f}}_d} \in {\mathbb{C}^{M_{t} \times 1}}$ is the beamforming vector, $\kappa _d$ and $b_d$ are delay and Doppler compensations,  respectively. ${\bf{s}}(t)$ is obtained by passing the digital signal through pulse shaping filter as
\begin{equation}
{\bf{s}}(t) = \sum\limits_{n =  - \infty }^\infty  {{\bf{s}}[n]p(t - nT)},
\label{eq21}
\end{equation}
where $p(t)$ is the pulse shaping filter impulse response. After passing through the linear time variant channel (LTV) channel, the received signal is\cite{b8} 
\begin{equation}
y(t) = \int_{ - \infty }^\infty  {\int_{ - \infty }^\infty  {{\bf{h}}{{(\nu ,\tau )}^H}} } {\bf{s}}(t - \tau ){e^{j2\pi \nu t}}d\tau d\nu  + w(t),
\label{eq3}
\end{equation}
where ${{\bf{h}}(\nu ,\tau )\in {\mathbb{C}^{M_{t} \times 1}}}$ represents the delay-Doppler spreading function and $w(t)$ is Additive White Gaussian noise (AWGN). In mmWave and THz frequency bands, the channel will exhibit sparsity. Specifically, the channel has ${L}$ paths, with the $p$th path having path delay ${\tau _p}$, Doppler shift ${\nu _p}$, and channel vector ${{\bf{h}}_p}$. ${\tau _p}$ and ${\nu _p}$ can be decomposed into integer and a decimal components, i.e., ${\tau _p} = {l_p}T = ({l_{pi}} + {l_{pf}})T,{\nu _p} = \frac{{{k_p}}}{{NMT}} = \frac{{({k_{pi}} + {k_{pf}})}}{{NMT}}$, where ${l_{pi}}$, ${l_{pf}}$, ${k_{pi}}$ and ${k_{pf}}$ represent the integer and fractional parts of the $p$th path delay and Doppler shift after sampling. As a result, the spreading function ${{\bf{h}}(\nu ,\tau )}$ can be written as
\begin{equation}
{\bf{h}}(\nu ,\tau ) = \sum\limits_{p = 0}^{L - 1} {{{\bf{h}}_p}\delta (\tau  - {\tau _p})} \delta (\nu  - {\nu _p}).
\label{eq4}
\end{equation}
By substituting (\ref{eq4}) into (\ref{eq3}), the received signal is 
\begin{equation}
y(t) = \sum\limits_{p = 0}^{L - 1} {{{\bf{h}}_p^H}{\bf{s}}(t - {\tau _p}){e^{j2\pi {\nu _p}t}}}  + w(t).
\end{equation}
The received signal after sampling with the sampling interval $T$ is
\begin{flalign}
\begin{split}
y[n] &= \sum\limits_{p = 0}^{L - 1} {{\bf{h}}_p^H\sum\limits_{m =  - \infty }^\infty  {{\bf{s}}[m]} p((n - m - {l_p})T)} {e^{j\frac{{2\pi {k_p}}}{{NM}}n}}\\
&+ w[n].
\label{eq99}
\end{split}&
\end{flalign}
Performing DZT yields $Y_{DD}[k,l]$ as
\begin{equation}
\label{eq98}
{Y_{DD}}[k,l] = \frac{1}{{\sqrt N }}\sum\limits_{c = 0}^{N - 1} {y[n]{e^{ - j2\pi \frac{c}{N}k}}}{|_{n = l + cM}}.
\end{equation}
Considering the occurrence of fractional Doppler shifts and fractional delays in practical communication scenarios, there are two implementation approaches for DDAM processing. The first approach is to align based on the physical paths. This method is suitable when the multi-paths are well resolved so that the parameters of each individual path are estimated. The second approach is to perform alignment based on delay-Doppler bins. This approach is suitable when the multi-paths cannot be well-resolved and it is difficult to accurately estimate the delay and Doppler shift of each path in the channel. The specific implementations and input-output relationships of these two approaches will be discussed next.
\subsubsection{Alignment based on path}\label{AA}
% \begin{equation}
% {\tau _p} = {\tau _{pi}} + {\tau _{pf}} = {l_p}T = ({l_{pi}} + {l_{pf}})T,\notag
% \end{equation}
% \begin{equation}
% {\nu _p} = {\nu _{pi}} + {\nu _{pf}} = \frac{{{k_p}}}{{KLT}} = \frac{{({k_{pi}} + {k_{pf}})}}{{KLT}},
% \label{eq20}
% \end{equation}
(\ref{eq99}) without noise can be written as
\begin{flalign}
\begin{split}
&\tilde y[n]= \sum\limits_{p = 0}^{L - 1} {{\bf{h}}_p^H\sum\limits_{m =  - \infty }^\infty  {{\bf{s}}[m]} p((n - m - {l_p})T)} {e^{j\frac{{2\pi {k_p}}}{{NM}}n}}\\
&=\sum\limits_{p = 0}^{L - 1} {{\bf{h}}_p^H\sum\limits_{m =  - \infty }^\infty  {{\bf{s}}[m]} p((n - m - {l_{{p_i}}})T - {l_{{p_f}}}T)} {e^{j\frac{{2\pi {k_p}}}{{NM}}n}}\\
&= \sum\limits_{p = 0}^{L - 1} {{\bf{h}}_p^H\sum\limits_{z =  - \infty }^\infty  {{\bf{s}}[n - z - {l_{{p_i}}}]} p(zT - {l_{{p_f}}}T)} {e^{j\frac{{2\pi {k_p}}}{{NM}}n}},
\label{eq22}
\end{split}&
\end{flalign}
where we have used change of variable $z = n - m - {l_{{p_i}}}$. 

This approach aligns $\{ p = 0,1,...,L-1\}$  physical paths, for which D in (\ref{eq16}) is set to $L$. By substituting (\ref{eq16}) into (\ref{eq22}), we have
\begin{equation}
\tilde y[n] = \sum\limits_{p = 0}^{L - 1} {\sum\limits_{p' = 0}^{L - 1} {\sum\limits_{z =  - \infty }^\infty  {} } } \notag
\end{equation}
\begin{equation}
{\bf{h}}_p^H{{\bf{f}}_{p'}}x[n - z - {l_{{p_i}}} - {\kappa _{p'}}]p(zT - {l_{pf}}T){\theta _{p,p'}},
\label{eq24}
\end{equation}
% \begin{equation}
% {\theta _{p,p'}} = 2\pi (({b_{p'}} + {k_{{p_i}}})n - ({b_{p'}}{l_{{p_i}}} + {k_{p{'_i}}}{l_{p{'_i}}})),\notag
% \end{equation}
where ${\theta _{p,p'}} = j2\pi \frac{{{k_p}n + {b_{p'}}(n - z - {l_{pi}})}}{{NM}}$. Let ${\kappa _p} + {l_{pi}} = {n_{\max }}$, ${b_p} + {k_{p}} = 0$ and ${\bf{h }}_p^H{{\bf{f}}_{p'}} = 0,$ $\forall p \ne p'$(\cite{b4} provides specific design of ${{\bf{f}}_p}$ such as ISI-maximal-ratio transmission (MRT) beamforming, i.e., ${{\bf{f}}_p} = \sqrt P {{\bf{h}}_p}/\left\| {{{\bf{h}}_p}} \right\|$, where $P$ denotes the available transmit power), we have
\begin{flalign}
\begin{split}
&\tilde y[n] = \sum\limits_{p = 0}^{L - 1} {\sum\limits_{z =  - \infty }^\infty  {{\bf{h}}_p^H{{\bf{f}}_p}p(zT - {l_{pf}}T)s[n - z - {n_{\max }}]{\theta _{p,p}}} } \\
& =\sum\limits_{p = 0}^{L - 1} {\overline {\bf{h}} _p^H{{\bf{f}}_p}p( - {l_{pf}}T)s[n - {n_{\max }}]}  \\
& + \sum\limits_{z \ne 0}^{} {\left( {\sum\limits_{p = 0}^{L - 1} {\overline {\bf{h}} _p^H{{\bf{f}}_p}p(zT - {l_{pf}}T)} {e^{j2\pi \frac{{{k_p}z}}{{NM}}}}} \right)s[n - z - {n_{\max }}]} ,
\end{split}&
\end{flalign}
where ${\overline {\bf{h}} _p} = {{\bf{h}}_p}{e^{j2\pi \frac{{{k_p}{l_{pi}}}}{{NM}}}}$. Then according to (\ref{eq98}), we can get ${Y_{DD}}[k,l]$ with noise as
\begin{flalign}
\begin{split}
&{Y_{DD}}[k,l] = \sum\limits_{p = 0}^{L - 1} {\overline {\bf{h}} _p^H{{\bf{f}}_p}p( - {l_{pf}}T){X_{DD}}[k,l - {n_{\max }}] + } \\
& \sum\limits_{z \ne 0}^{} {\left( {\sum\limits_{p = 0}^{L - 1} {\overline {\bf{h}} _p^H{{\bf{f}}_p}p(zT - {l_{pf}}T)} {e^{j2\pi \frac{{{k_p}z}}{{NM}}}}} \right)} {X_{DD}}[k,l - {n_{\max }} - z]\\
&+ {W}[k,l],
\end{split}&
\end{flalign}
where ${W}[k,l]$ is the noise in delay-Doppler domain with a power spectral density of ${N_0}$. The first term is the desired signal. The second term is the interference. Due to the existence of fractional multi-path delays, ISI still exists after compensating the integer delays. As $z$ increases, the energy of the interference will quickly decay. Therefore, it can be approximated by considering only ${N_i}$ interference signals around the aligned symbols. At this point, the delay spread of the signal is also approximately reduced to $\tau {'_{spread}} = 2{N_i}T$, and the Doppler spread is reduced to 0. The signal-to-interference-plus-noise ratio (SINR) of the DDAM-OTFS system with path-based alignment is
\begin{equation}
{\gamma _1} = \frac{{{{\left| {\sum\limits_{p = 0}^{L - 1} {\overline {\bf{h}} _p^H{{\bf{f}}_p}p( - {l_{pf}}T)} } \right|}^2}}}{{\sum\limits_{z \ne 0}^{} {{{\left| {\sum\limits_{p = 0}^{L - 1} {\overline {\bf{h}} _p^H{{\bf{f}}_p}p(zT - {l_{pf}}T){e^{j2\pi \frac{{{k_p}z}}{{NM}}}}} } \right|}^2}}  + {N_0}}}.
\end{equation}

% Here is a specific example to illustrate this result: Transmit an OTFS symbol only map at the origin, pass through a 3-path channel to obtain ${Y_{DD}}[k,l]$. Figure AAAA shows the result without DDAM processing, where it can be seen that the signal generates corresponding interference over a large range around it, with significant delay and Doppler spreads. Figure BBBBBB shows the result with the addition of DDAM processing, where it can be observed that the channel only has relatively small interference on the delay dimension around the symbols, which decays rapidly, and disappears completely in the Doppler dimension. The delay and Doppler spreads of the channel are greatly reduced.
\subsubsection{Alignment based on bin}\label{BB}
Rewrite (\ref{eq99}) as
\begin{equation}
y[n] =\sum\limits_{m =  - \infty }^\infty  {\sum\limits_{k =  - \infty }^\infty  {{\bf{s}}[n - m]} } {\bf{h}}_{bin}^H[k,m]{e^{j2\pi \frac{kn}{{MN}}}},
\label{eq25}
\end{equation}
where ${{\bf{h}}_{bin}}[k,m]$ is the sum of all channel responses at 
the delay-Doppler bin [$k, m$], given by
\begin{flalign}
{{\bf{h}}_{bin}}[k,m] = \sum\limits_{p = 0}^{L-1} {{{\bf{h}}_p}} p((m - {l_p})T)\mathcal{G} (k - {k_p}),
\end{flalign}
where ${\mathcal{G} (k - {k_p}) = \frac{{{e^{ - j2\pi (k - {k_p})}} - 1}}{{{e^{ - j\frac{{2\pi }}{N}(k - {k_p})}} - 1}}}$\cite{b369}. 

In the delay-Doppler domain, the channel response in general has more than $L$ non-zero bins due to the existence of fractional delays and Doppler shifts. Here, we only align specific bins among them. The bin that meets the following conditions are chosen to be aligned: (1) ${\left\| {{{\bf{h}}_{bin}}[k,m]} \right\|^2} \ge C\max \left\{ {{{\left\| {{{\bf{h}}_{bin}}[k,m]} \right\|}^2}} \right\}$, where \textit{C} $<$ 1 is certain threshold. (2) ${\left\| {{{\bf{h}}_{bin}}[k,m]} \right\|^2} \ge \max \{ {\left\| {{{\bf{h}}_{bin}}[k,m + 1]} \right\|^2},{\left\| {{{\bf{h}}_{bin}}[k + 1,m]} \right\|^2},{\left\| {{{\bf{h}}_{bin}}[k,m - 1]} \right\|^2},$ ${\left\| {{{\bf{h}}_{bin}}[k - 1,m]} \right\|^2}\} $.
The bins selected in this way are generally the nearest integer bins associated with each physical path. Denoting the index of each bin as \textit{a}, the position of each bin is expressed as $({k_\alpha },{l_\alpha })$, and the set of bins that meet the threshold condition is represented by $\Phi$, other bins correspond to energy leakage caused by fractional delays and Doppler shifts. They are located around the selected bins, forming $L$ clusters, and will be shifted along with the selected bins in delay and Doppler compensation since the same cluster of bins corresponds to one physical path \cite{b381}. The set of ${L'}$ selected bins is: $\alpha  \in \Omega  = \{ ({k_{{g}}},{l_{{g}}})|g = 0,...,L' - 1\} $. In this approach, $D$ in (\ref{eq16}) is set to $L'$ and the signal in (\ref{eq16}) can be represented as
\begin{equation}
{\bf{s}}[n] = \sum\limits_{g = 0}^{L' - 1} {{{\bf{f}}_g}x[n - {\kappa _g}]{e^{j2\pi \frac{{{b_g}n}}{{MN}}}}} .
\label{eq97}
\end{equation}
Substitute (\ref{eq97}) into (\ref{eq25}), the received signal is
\begin{flalign}
\begin{split}
&y[n] = \sum\limits_{g = 0}^{L' - 1} {\sum\limits_{g' = 0}^{L' - 1} {{\bf{h}}_{bin}^H[{k_g},{l_g}]} } {{\bf{f}}_{g'}}x[n - {l_g} - {\kappa _{g'}}]{e^{\frac{{j2\pi {\mu _{g,g'}}}}{{MN}}}}\\
& + \sum\limits_{\alpha  \notin \Omega } {\sum\limits_{g' = 0}^{L' - 1} {{\bf{h}}_{bin}^H[{k_\alpha },{l_\alpha }]{{\bf{f}}_{g'}}x[n - {l_\alpha } - {\kappa _{g'}}]{e^{\frac{{j2\pi {\mu _{\alpha ,g'}}}}{{MN}}}}} } \\
&+ w[n],
\end{split}&
\label{eq26}
\end{flalign}
where ${\mu _{g,g'}} = {b_{g'}}(n - {l_g}) + {k_g}n$. By letting ${\kappa _g} + {l_{g}} = {n_{\max }},{b_g} + {k_{g}} = 0$ and ${\bf{h}}_{bin}^H[{k_{{g}}},{l_{{g}}}]{{\bf{f}}_{g'}} = 0,$ $\forall g'\neq g$, (\ref{eq26}) reduces to 
\begin{flalign}
\begin{split}
&y[n] = (\sum\limits_{g = 0}^{L' - 1} {\overline {\bf{h}} _{bin}^H[{k_g},{l_g}]} {{\bf{f}}_g})x[n - {n_{\max }}]\\
& + \sum\limits_{\alpha  \notin \Omega } {\sum\limits_{g' = 0}^{L' - 1} {{\bf{h}}_{bin}^H[{k_\alpha },{l_\alpha }]{{\bf{f}}_{g'}}x[n - {l_\alpha } - {\kappa _{g'}}]{e^{\frac{{j2\pi {\mu _{\alpha ,g'}}}}{{MN}}}}} }  + w[n],
\end{split}&
\label{eq27}
\end{flalign}
where $\overline {\bf{h}} _{bin}^H[{k_g},{l_g}] = {\bf{h}}_{bin}^H[{k_g},{l_g}]{e^{j2\pi \frac{{{k_g}{l_g}}}{{MN}}}}$. Taking DZT on both sides of the equation, we get
\begin{flalign}
\begin{split}
&{Y_{DD}}[k,l] = (\sum\limits_{g = 0}^{L' - 1} {\overline {\bf{h}} _{bin}^H[{k_g},{l_g}]{{\bf{f}}_g}} ){X_{DD}}[k,l - {n_{\max }}] + \\
&\sum\limits_{\alpha  \notin \Omega } {\sum\limits_{g' = 0}^{Lo - 1} {{\bf{h}}_{bin}^H[{k_\alpha },{l_\alpha }]{{\bf{f}}_{g'}}} } {X_{DD}}[k - {k_\alpha } - {b_{g'}},l - {l_\alpha } - {\kappa _{g'}}]{\vartheta}\\
&+ W[k,l],
\end{split}&
\label{eq28}
\end{flalign}
where $\vartheta  = e\frac{{j2\pi }}{{MN}}[({k_\alpha } - {k_{g'}})({l_\alpha } + {\kappa _{g'}}) + {k_{g'}}{l_\alpha }]$. To derive the SINR with residual ISI, it is necessary to re-organize the interference terms in the second part. Let $z = {l_\alpha } - l _{g'},$$i = {k_\alpha } - {k_{g'}}$, we have
\begin{flalign}
\begin{split}
&{Y_{DD}}[k,l] = (\sum\limits_{g = 0}^{L' - 1} {\overline {\bf{h}} _{bin}^H[{k_g},{l_g}]{{\bf{f}}_g}} ){X_{DD}}[k,l - {n_{\max }}] \\
&  + \sum\limits_{(i,z) \ne (0,0)} {{X_{DD}}[k - i,l - {n_{\max }} - z]}  \cdot {\rm{ }} \\
&\left( {\sum\limits_{g' = 0}^{L' - 1} {{\bf{h}}_{bin}^H[{k_{g'}} + i,{l_{g'}} + z]{{\bf{f}}_{g'}}\vartheta '} } \right) + W[k,l],\\
\end{split}&
\label{eq29}
\end{flalign}
where $\vartheta ' = e\frac{{j2\pi }}{{MN}}[i({n_{\max }} + z) + {k_{g'}}(z + {l_{g'}})]$. Compared with the first approach, the second term here introduces additional interference on the Doppler axis. As $i$ and $z$ increase, the interference energy will rapidly decay. Therefore, it can be approximated by considering only ${N_i}$ interfering signals near the aligned symbol on the delay axis and ${K_i}$ on the Doppler axis. The delay and Doppler spreads are approximately reduced to $\tau {'_{spread}} = 2{N_i}T$ and $\nu {'_{spread}} = \frac{{2{K_i}}}{{MNT}}$, respectively. The SINR is
\begin{equation}
{\gamma _2} = \frac{{{{\left| {\sum\limits_{g = 0}^{L' - 1} {\overline {\bf{h}} _{bin}^H[{k_g},{l_g}]{{\bf{f}}_g}} } \right|}^2}}}{{\sum\limits_{(i,z) \ne (0,0)} {{{\left| {\sum\limits_{g' = 0}^{L' - 1} {{\bf{h}}_{bin}^H[{k_{g'}} + i,{l_{g'}} + z]{{\bf{f}}_{g'}}\vartheta '} } \right|}^2} + {N_0}} }}.
\label{eq30}
\end{equation}

\begin{figure}[H]
\centerline{\includegraphics[width=0.5\textwidth]{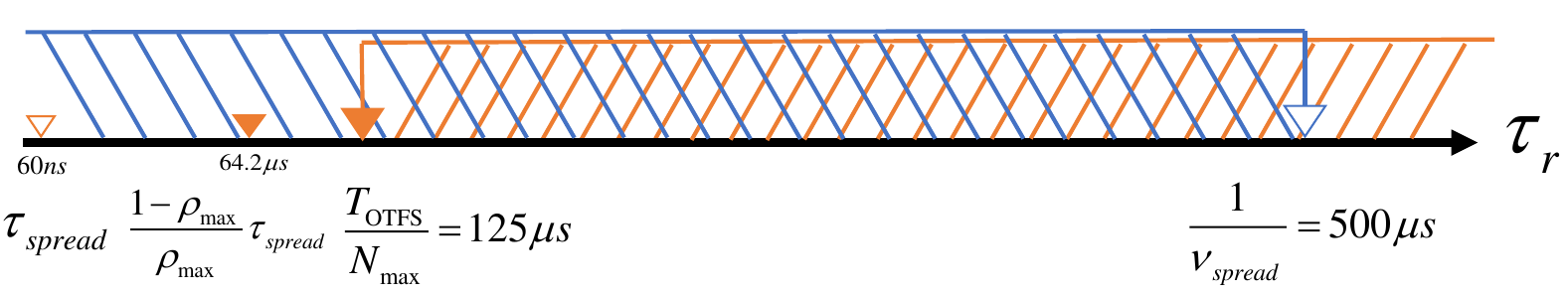}}
\caption{ The constraints on OTFS period parameters after DDAM.}
\label{fig9}
\end{figure}

% Here is a simulation between the equivalent channels before and after DDAM processing in a three-path channel. The x and y axes represent the delay and doppler dimensions, respectively, while the z axis represents the channel response magnitude：
% \begin{figure}[H]
% \centerline{\includegraphics[width=0.5\textwidth]{path.pdf}}
% \caption{ The equivalent channel before DDAM.}
% \label{fig999}
% \end{figure}
% \begin{figure}[H]
% \centerline{\includegraphics[width=0.5\textwidth]{bin.pdf}}
% \caption{ The equivalent channel after DDAM.}
% \label{fig99}
% \end{figure}
After DDAM processing, the selection of OTFS period parameters will become more flexible. For the example shown in Fig. \ref{fig6}, after DDAM processing, the delay and Doppler spreads of the channel are reduced to 60 $ns$ and 2 $kHz$, as shown in Fig. \ref{fig9}, allowing previously infeasible cases to become feasible, i.e.,  $125\,\mu s \leq {\tau _r}{\rm{  < 500}}\,\mu {\rm{s}}$.
\section{NUMERICAL RESULTS}
\label{section:C} %这里是章节的标签，引用时需要
In this section, we provide simulation results to evaluate the performance of the proposed DDAM-OTFS. We first evaluate the average SE of DDAM-OTFS under the scenario where the UE is moving at a high speed. Then we evaluate the performance of DDAM-OTFS system in terms of PAPR.
The basic parameters for the simulations are set as follows. The carrier frequency ${f_c}$ is 28 GHz, the total bandwidth $B$ is 64 MHz, and the noise power spectral density ${N_0}$ is $-174$ dBm/Hz. The transmitter is equipped with a uniform linear array (ULA) with half-wavelength spacing between adjacent elements. The number of temporal-resolvable multi-paths with non-negligible path strength is ${L} = 5$, with path delays uniformly distributed in $[0,{\tau _{\max }}]$, where ${\tau _{\max }} = 500\,ns$. The UE's maximum speed is 300 km/h. The corresponding Doppler spread is ${v_{spread}}=15.57\,kHz$. The frame duration is ${T_{\rm{OTFS}}} = 1\,ms$, with each  frame of the original OTFS consisting of 512 subcarriers and 128 time slots. The MRT beamforming is applied to OTFS in the TF domain \cite{b10}. 

\begin{figure}[H]
\centerline{\includegraphics[width=0.48\textwidth]{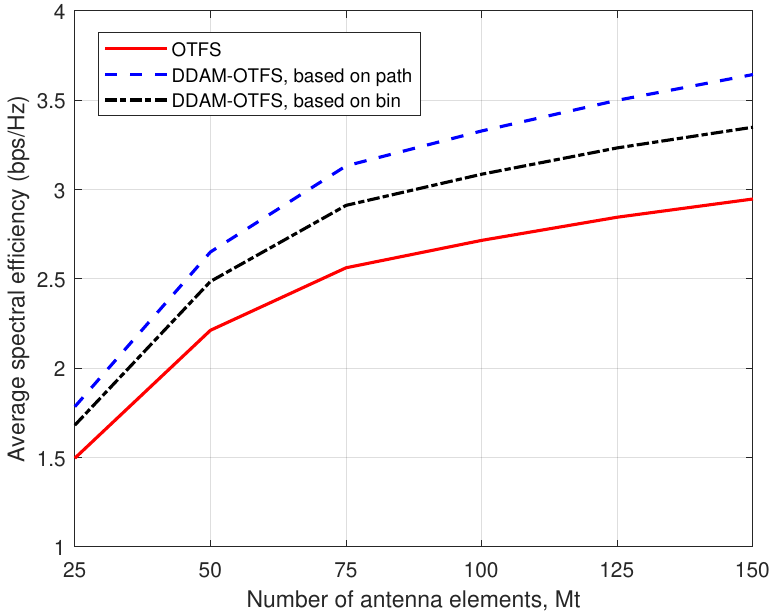}}
\caption{  SE versus the number of transmit antennas.}
\label{fig10}
\end{figure}

Fig. \ref{fig10} compares the SE of conventional OTFS, the proposed path-based and bin-based DDAM-OTFS, where the CP is inserted per time slot. The transmit power is 30 dBm. In this scenario, the CP overhead of DDAM-OTFS is significantly reduced due to the reduced channel delay spread.
The CP overhead for OTFS is given by $\frac{{B{\tau _{spread}}}}{{M + B{\tau _{spread}}}}{\rm{ }} \approx {\rm{5}}{\rm{.88\% }}$, and after processing with DDAM-OTFS, the reduced delay spread becomes $\tau {'_{spread}} = 2{N_i}T = 62.5\,ns$, resulting in a reduced CP overhead of $0.735\%$. The results in Fig. \ref{fig10} show that for different numbers of antennas, both implementations of DDAM-OTFS proposed in this paper achieve higher average SE than traditional OTFS. This is expected due to the saving of CP overhead with our proposed DDAM-OTFS.

\begin{figure}[H]
\centerline{\includegraphics[width=0.48\textwidth]{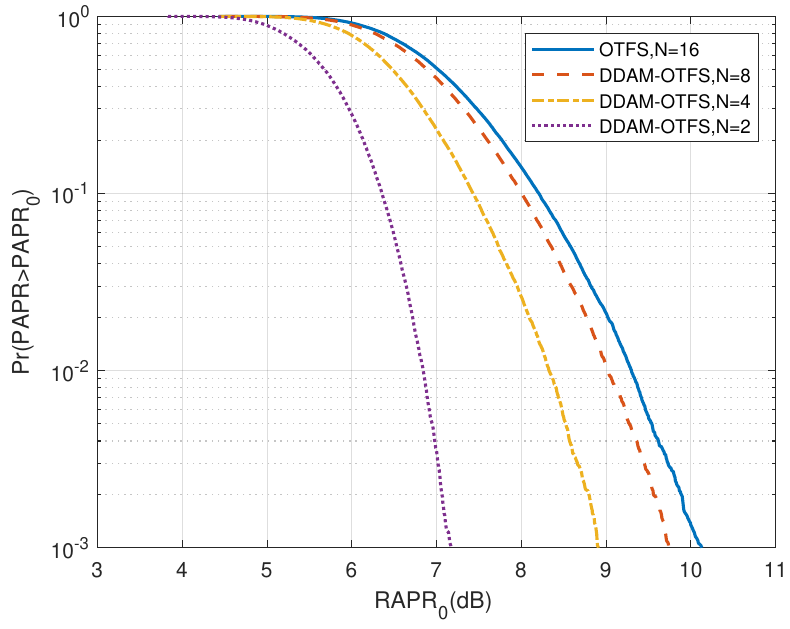}}
\caption{ PAPR for the proposed DDAM-OTFS and the traditional OTFS.}
\label{fig11}
\end{figure}

Fig. \ref{fig11} compares the PAPR of the original OTFS and the proposed DDAM-OTFS schemes. The PAPR of the OTFS system is related to the number of time slots $N$ in an OTFS frame. Since $\frac{{{T_{OTFS}}}}{{{N}}} < \frac{1}{{{\nu _{spread}}}}$, the minimum value of $N$ for OTFS without DDAM processing is 16. However, DDAM-OTFS reduces the minimum requirement of $N$ by shortening the Doppler spread, allowing for $N = 2$, $4$, 8. This reduction in $N$ leads to the reduced PAPR. It can be observed that DDAM-OTFS exhibits significantly improved performance compared to traditional OTFS. It is also appealing since it helps mitigate the practical issue of PAPR for OTFS transmission, which eases the requirement of linear region of
power amplifiers compared to the conventional OTFS.
\section{CONCLUSION}
\label{section:D} %这里是章节的标签，引用时需要
This paper proposed a novel transmission scheme called DDAM-OTFS. We first discussed the CP overhead and PAPR of OTFS in relation to the constraints imposed by channel delay and Doppler spreads, elucidating how the introduction of DDAM relaxes these constraints by reducing channel delay and Doppler spreads. Subsequently, the paper proposed two implementations for DDAM-OTFS, catering to different scenarios, namely
path-based alignment and bin-based alignment, along with the corresponding input-output relationships and SINR performance for both approaches. Finally, through a series of simulations, we validated the enhancement in average SE and PAPR performance of the proposed DDAM-OTFS compared to traditional OTFS.

\end{document}